\documentclass[aps, prb, showpacs, twocolumn, amsmath, letterpaper]{revtex4-2}

\usepackage{graphics}
\usepackage{graphicx}
\usepackage{amssymb}
\usepackage{bm}
\usepackage{xcolor}
\usepackage{textcomp}
\usepackage{siunitx}

\begin{document}

\title{Tunable Giant Anomalous Hall in a Kondo Lattice Ferromagnet UBiTe}

\author{Qiaozhi Xu$^{1}$, Hasan Siddiquee$^{1}$, Shannon Gould$^{1}$, Jiahui Althena Zhu$^{1}$, David Alonso Martinez$^{1}$, Christopher Broyles$^{1}$, Guangxin Ni$^{2,3}$, Tai Kong$^{4}$ \& Sheng Ran$^{1}$}

\affiliation{$^1$ Department of Physics, Washington University in St. Louis, St. Louis, MO 63130, USA
\\$^2$ Department of Physics, Florida State University, Tallahassee, FL 32306, USA
\\$^3$ National High Magnetic Field Laboratory, Tallahassee, FL 32310, USA
\\$^4$ Department of Physics, University of Arizona, Tucson, AZ 85721, USA
}

\date{\today}

\begin{abstract}

Kondo lattice systems are recognized for potentially hosting a variety of rich topological phases. Several pioneering studies have demonstrated significant anomalous Hall and anomalous Nernst effects in these systems, attributed to the Berry curvature of the hybridization bands. In this study, we investigate UBiTe, a ferromagnetic Kondo lattice system. Our findings reveal that the intrinsic contribution to the anomalous Hall conductivity is closely tied to the Kondo coherence temperature. Moreover, we demonstrate that slight shifts in the Fermi level across three different samples significantly influence this intrinsic contribution, aligning with the Berry curvature localized within the narrow hybridization bands. This provides a stark contrast to the less pronounced sensitivity observed in weakly correlated Weyl semimetals, underscoring the distinctive electronic properties of Kondo lattice systems. The anomalous Hall conductivity of one sample ranks among the highest reported for topological magnetic materials. 

\end{abstract}
\maketitle
\section{Introduction}

Strongly correlated materials are crucial for exploring topological properties for the following reasons. First, due to strong Coulomb interactions, it is highly challenging to calculate and predict their topological properties, rendering these systems particularly significant for advancing theoretical physics. Second, strong correlations are known to intertwine different degrees of freedom and give rise to exotic excitations, suggesting that strongly correlated materials may host new topological phases that have no non-interacting counterpart, such as the fractional quantum Hall state~\cite{Stormer1999,Park2023,Lu2024}. Kondo lattice systems are a prototype of strongly correlated materials that might host rich topological phases. Kondo entanglement between the localized magnetic moments and the conduction electrons leads to the formation of narrow hybridization bands near the Fermi level. If these hybridized bands exhibit nontrivial topology, it guarantees the presence of strongly correlated topological phases, as these renormalized bands do not exist at high temperatures and emerge solely due to correlation effects. In addition, since these bands are situated close to the Fermi level, their nontrivial topology will have a profound impact on the material's physical properties~\cite{Alex2013}.

Starting from the original proposal that Kondo insulators can be topological~\cite{Dzero2016}, the field has been enriched by the discoveries of other Kondo topological materials, such as Kondo Weyl semimetals~\cite{Lai2018, Ivanov2019,Grefe2020,Dzsaber2021,Chen2022}. Several candidates for Kondo topological materials demonstrate a substantial anomalous Hall or anomalous Nernst effect~\cite{Asaba2022,Siddiquee2023,Li2024}. An open question remains: Are these significant Berry curvature effects related to the hybridization between $f$-electrons and conduction electrons? 

We have recently studied USbTe~\cite{Siddiquee2023}, a Kondo lattice ferromagnet. We observed large anomalous Hall conductivity (AHC), with a sign change upon cooling the system, which cannot be attributed to skew scattering, as the direction of the magnetic order does not change. Instead, a combination of our scaling analysis and ARPES measurements indicates that the anomalous Hall conductivity at low temperatures is dominated by the non-trivial Berry curvature arising from the Kondo-hybridized bands. This is further supported by our first-principles calculations, which show that the $f$ bands host a large number of Weyl nodes that give rise to the large AHC.

In this study, we focus on UBiTe~\cite{Hulliger1968}, another Kondo lattice ferromagnet with the same crystal structure as USbTe. This investigation aims to further unravel the relationship between Kondo hybridization and topological properties. Our results reveal a strong correlation between the intrinsic contributions to AHC and the Kondo coherence temperature. In addition, we found that slight variations in the Fermi level significantly affect the anomalous Hall conductivity in this material. This is consistent with the presence of Berry curvature in narrow bands arising from Kondo hybridization. These findings establish UBiTe as a promising platform for exploring the interplay between magnetism, topological properties, and strong correlations.


\section{Methods}

Single crystals of UBiTe were synthesized by the molten flux method using excess BiTe as flux. The elements were mixed in the ratio U:Te:Bi
= 2:9:9. The crucible containing the starting elements was sealed in a fused silica ampule and then heated up to 1000~$^{\circ}$C in a box furnace. The crystals grew as the temperature was reduced to 600~$^{\circ}$C over 100 hours, after which the ampule was quickly removed from the furnace and the flux was decanted in a centrifuge. The crystal structure was determined by powder X-ray diffraction using a Rigaku X-ray diffractometer with Cu-$K_{\alpha}$ radiation. A small amount of BiTe impurity peaks were detected, likely due to residual flux on the surface. Electrical transport measurements were performed in a Quantum Design physical property measurement system (PPMS). An electric current was applied along the $a$ axis, and Hall voltage was measured in the perpendicular direction. Positive and negative magnetic fields were applied to antisymmetrize the Hall signal. Magnetization measurements were performed in a Quantum Design PPMS with a VSM option. Specific heat measurements were also performed in a Quantum Design PPMS. Energy-dispersive X-ray spectroscopy measurements were performed using a Thermo Fisher Quattro environmental scanning electron microscope.

\section{Results}

UBiTe crystallizes in a nonsymmorphic crystal structure with space group 129 (P4/nmm)~\cite{Hulliger1968}, the same as USbTe. Detailed physical properties have not been studied before. U and Te atoms each form planes with mirror and screw nonsymmorphic symmetries. Similar to USbTe, UBiTe exhibits typical behaviors of a ferromagnetic Kondo lattice system, with a Curie temperature of $T_{c}$ = 106.5~K $\pm$ 1.5~K, evidenced by resistivity, magnetization, and specific heat measurements (Fig.~1). Determination of the transition temperature can be found in the Supplemental Material~\cite{supp1}(see also references ~\cite{Fisher1968,Ran2011,Zhong2003} therein). The magnetization shows large anisotropy in the ferromagnetic state with dominantly out-of-plane magnetic moment. The high-temperature magnetization follows the Curie-Weiss law with an effective magnetic moment $\mu_{eff}$ = 3.2~$\mu_{B}$/U, while the ordered moment is $\mu_{s}$ = 2.2~$\mu_{B}$/U along the $c$ axis. 

\begin{figure}[t!]
    \includegraphics[width=1\linewidth]{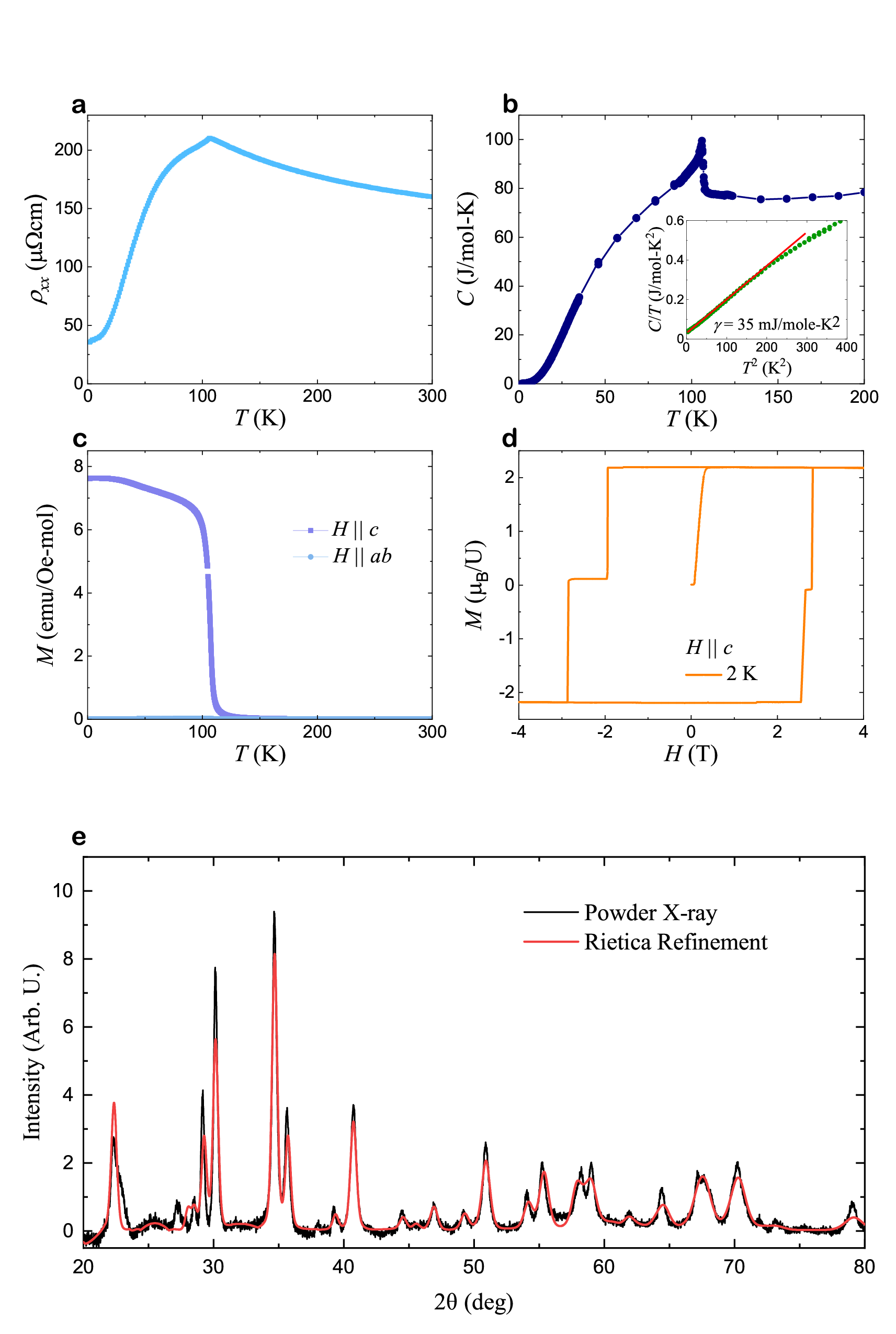}
    \caption{(a) Temperature dependence of the longitudinal electric resistivity $\rho_{xx}$ in zero-field for UBiTe single crystal. Current is applied along $a$ axis. (b) Temperature dependence of specific heat $C$ of UBiTe single crystal. Inset: $C/T$ as a function of $T^2$ showing the Sommerfeld coefficient as intercept. (c) Magnetization $M$ of UBiTe single crystal with magnetic field of 0.1~T applied along the $c$ axis and $ab$ plane. (d) Magnetization $M$ of UBiTe single crystal as a function of magnetic field at 2~K. (e) ) The powder x-ray diffraction profile of UBiTe.} 
\end{figure}

\begin{figure*}[ht!]
    \includegraphics[width=1\linewidth]{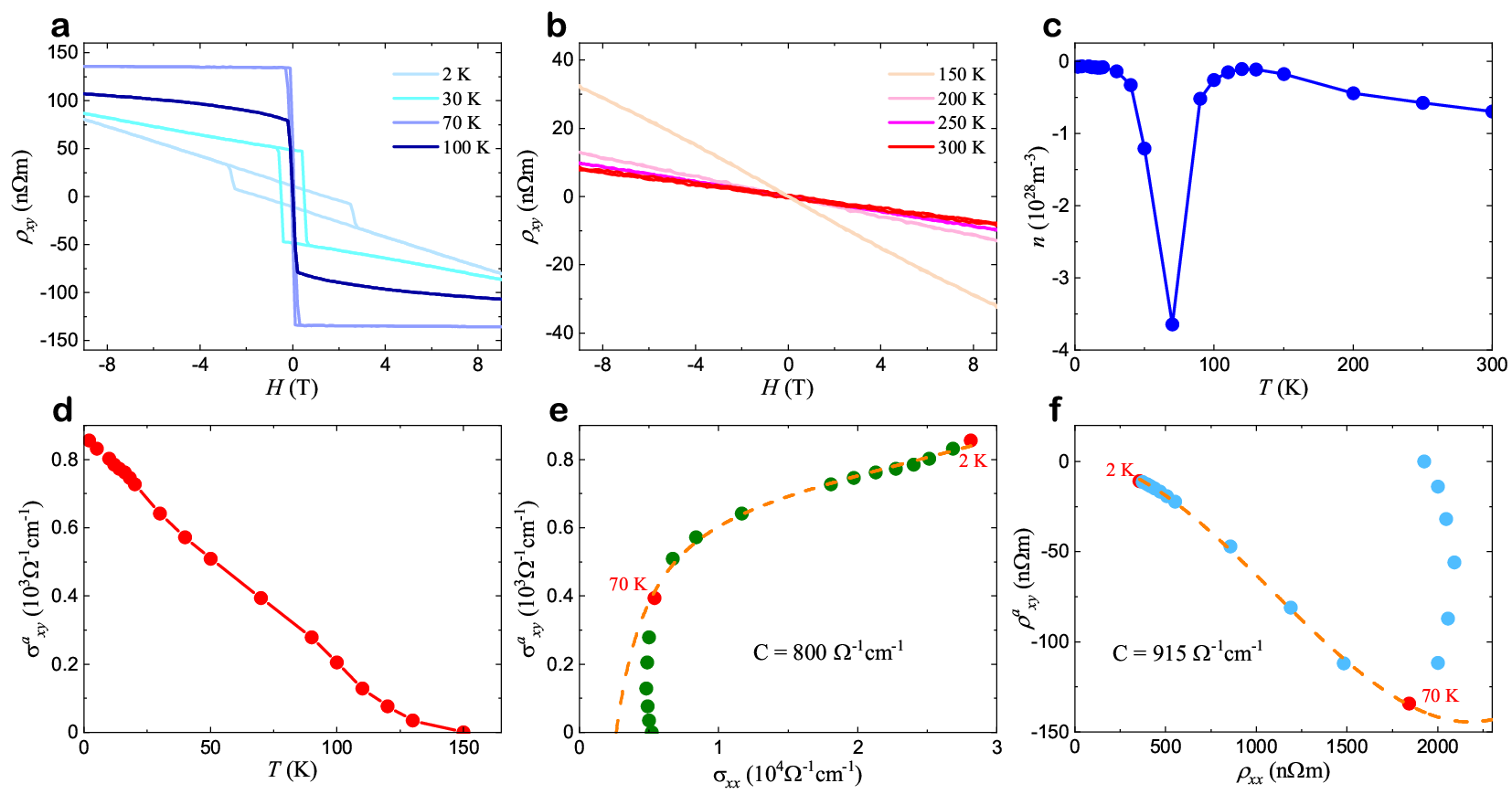}
    \caption{Anomalous Hall effect of sample S1. Magnetic field is applied along $c$ axis and electric current is along $a$ axis. (a) Magnetic field dependence of the Hall resistivity $\rho_{xy}$ at various temperatures below $T_c$. (b) Magnetic field dependence of the Hall resistivity $\rho_{xy}$ at various temperatures above $T_c$ showing linear slopes. (c) Carrier density extrapolated from the slop of the Hall resistivity as a function of temperature. Negative numbers are used to indicate that the carriers remain to be electrons for the whole temperature range. (d) Temperature dependence of anomalous Hall conductivity $\sigma^{a}_{xy}$ in zero magnetic field. (e) $\sigma^{a}_{xy}$ as a function of $\sigma_{xx}$ and the fit to $\sigma^a_{xy} = a\sigma^2_{xx} + b/\sigma_{xx} + c$ to extract the intrinsic contribution. (f) $\rho^{a}_{xy}$ as a function of $\rho_{xx}$ and the fit to $\rho^a_{xy} = -a - b\rho_{xx}^3 - c\rho_{xx}^2$ to extract the intrinsic contribution.}  
\end{figure*}

The Kondo lattice behavior is evidenced in resistivity data. The high-temperature resistivity shows a negative slope, well described by $-$ln$T$ dependence, due to the paramagnetic moments in the presence of single-ion Kondo hybridization with the conduction band. Below $T_{c}$, marked by the kink, resistivity decreases with decreasing temperature. However, the temperature dependence of resistivity cannot be described by scattering due to the ferromagnetic magnons, which gives rise to modified exponential behavior, such as $(k_BT/\Delta)^{3/2}Exp(\Delta/k_BT)$~\cite{Goodings1963}. Instead, $\rho_{xx}(T)$ shows a pronounced convex shape, not typical for metallic ferromagnets~\cite{Goodings1963}. This convex shape likely results from the combination of two effects: the ferromagnetic order, which leads to a decrease in resistivity below $T_{c}$, and the onset of Kondo coherence, which produces a broad maximum followed by a sharp drop in resistivity. The convex shape in resistivity is also observed in USbTe in a similar temperature range, coinciding with the emergence of the hybridization band as observed in ARPES measurements~\cite{Siddiquee2023}. This observation provides further indication that the convex shape in resistivity is a signature of the onset of Kondo coherence. Given the similarities between USbTe and UBiTe in physical properties including electric transport, magnetization and specific heat, we attribute the convex curvature in UBiTe to the onset of Kondo coherence as well, around 70~K. Unlike USbTe, which exhibits an upturn in resistivity at low temperatures, the resistivity of UBiTe continues to decrease across the entire temperature range below the Kondo coherence temperature.

The Sommerfeld coefficient of the electronic contribution to the specific heat is $\gamma_0 = 35$~mJ/mol-K$^2$ (inset to Fig.1b). At first glance, $\gamma_0$ does not seem to indicate very strong correlation. However, note that the system is already in a ferromagnetic state. In most other Kondo lattice systems, magnetic order, if it ever occurs, typically develops below the Kondo coherence temperature. In those cases, strong correlation is indicated by the $\gamma$ extracted above the magnetic order, and $\gamma_0$ at low temperature is typically much smaller~\cite{Jang2019,Posey2024}. Compared to those systems, a $\gamma_0$ of $35$~mJ/mol-K$^2$ for a magnetically ordered system is very large, indicating strong correlation in UBiTe.

Next, we focus on the anomalous Hall effect. Figure 2 shows the data for sample S1. Across the entire temperature range, the normal Hall effect exhibits negative linear slopes up to a magnetic field of 9~T, indicating electron carriers from a single band~\cite{Hurd1972,Coleman2007}. The carrier density extracted from a single-band Hall model shows a pronounced peak at 70~K, which is consistent with the onset of Kondo coherence. Below the Kondo coherence temperature, hybridization between the conduction electrons and the $f$-electrons leads to the formation of a hybridization gap. Although UBiTe is not a Kondo insulator and the Fermi level does not necessarily fall within this gap, the hybridization gap likely resides near the Fermi level and may partially overlap with states that contribute to electrical conduction, potentially reducing the carrier density.


Below $T_c$, on top of the normal Hall effect, there is a rectangular hysteresis loop with very sharp switching, indicating an anomalous Hall signal. The coercive field increases with decreasing temperature, reaching a value of 2~T at 2~K. The anomalous Hall conductivity $\sigma^a_{xy}$ as a function of temperature is summarized in Fig. 2d. The anomalous Hall effect can arise from two different mechanisms: extrinsic processes due to scattering effects, and an intrinsic mechanism connected to the Berry curvature~\cite{Nagaosa2010,Zhong2003}. Scattering effects depend on the scattering rate, leading to a strong dependence of $\sigma_{xy}$ on the longitudinal conductivity and temperature, while $\sigma_{xy}$ originating from the intrinsic mechanism is usually scattering-independent and, therefore, independent of $\sigma_{xx}$ and temperature~\cite{Liu2018,Wang2018,Huang2021}. The temperature dependence of $\sigma^a_{xy}$, as shown in Fig. 2d, seems to indicate a dominant scattering mechanism.

However, in a real material, $\sigma^a_{xy}$ can have contributions from both mechanisms. Separating the intrinsic anomalous Hall conductivity (AHC) from the extrinsic contributions is a significant challenge. The widely used scaling analysis is very simplified and may not apply to all ferromagnetic systems. It has been proposed that the following relation can capture both scattering and intrinsic effects well~\cite{Jones2022}: $\sigma^a_{xy} = a\sigma^2_{xx} + b/\sigma_{xx} + c$, where the first two terms represent the scattering effect, and the constant term represents the intrinsic contribution. Alternatively, the following relation can be used for anomalous Hall resistivity: $\rho^a_{xy} = -a - b\rho_{xx}^3 - c\rho_{xx}^2$. We applied these relations to our $\rho^a_{xy}$ and $\sigma^a_{xy}$ data, and both fit very well below 70~K. The intrinsic contributions obtained from both fittings are similar, with values of 800/$\Omega$-cm from $\sigma^a_{xy}$ and 915/$\Omega$-cm from $\rho^a_{xy}$. These fittings are only valid below 70~K, the Kondo coherence temperature. Deviations from the fitting models become pronounced above 70~K, especially in the $\rho^a_{xy}$ versus $\rho_{xx}$ data, which exhibits a distinct minimum at 70~K, dividing the dataset into two segments.

The alignment of the intrinsic contributions to AHC with the Kondo coherence temperature suggests a strong correlation between Berry curvature effects and the flat band associated with Kondo hybridization. In USbTe, through a combination of ARPES and transport measurements, we have similarly demonstrated this correlation~\cite{Siddiquee2023}. These hybridization bands are located within a narrow energy range close to the Fermi level. One might anticipate that even a small shift in the Fermi level could lead to significant changes in the Berry curvature of the flat band~\cite{Dirac1926,brab2024}, thereby influencing the anomalous Hall effect. Indeed, both the normal and anomalous Hall effects of UBiTe show significant sample variation, as discussed below.

\begin{figure*}[ht!]
    \includegraphics[width=1\linewidth]{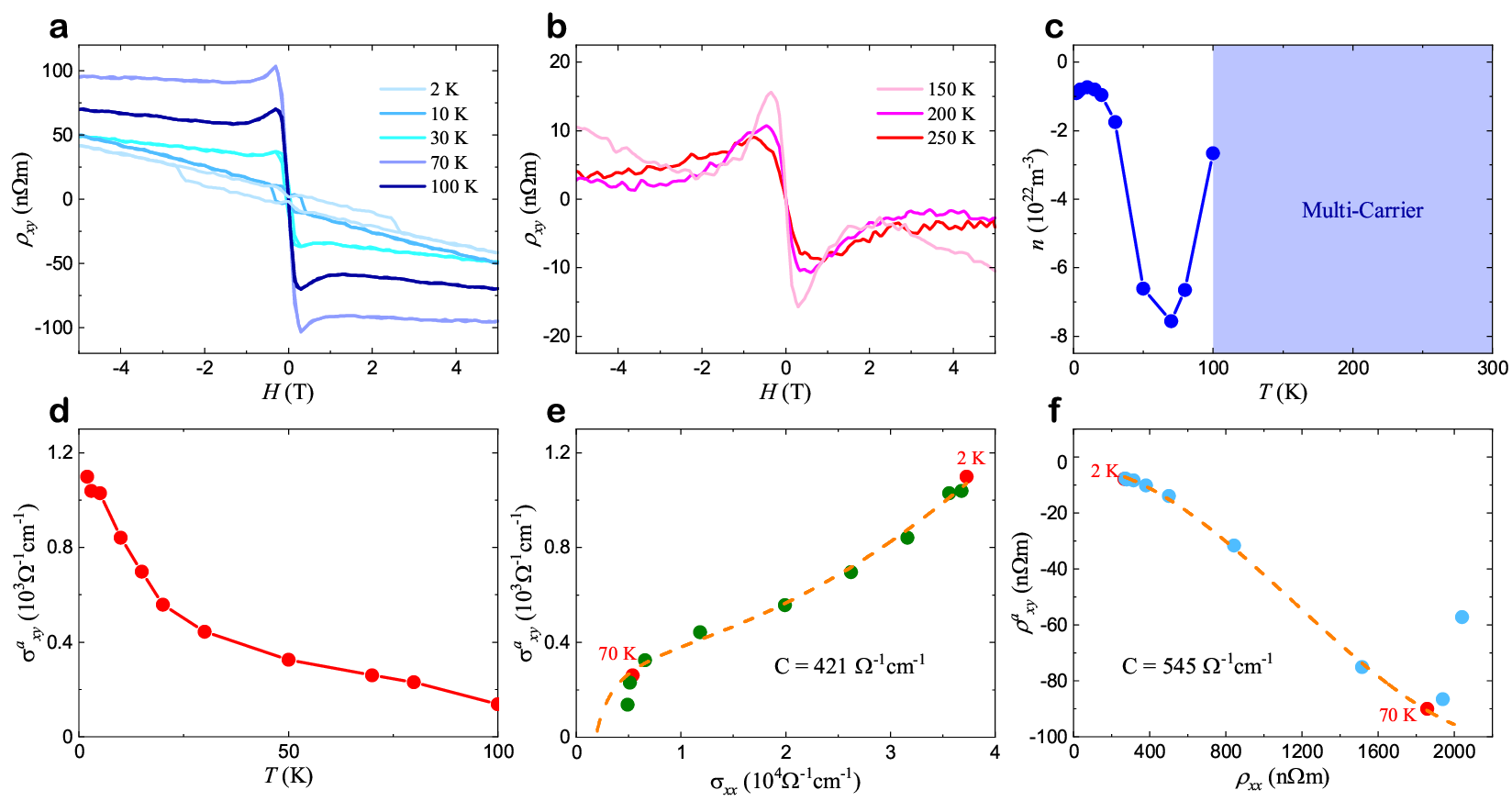}
    \caption{Anomalous Hall effect of sample S2. Magnetic field is applied along $c$ axis and electric current is along $a$ axis. (a) Magnetic field dependence of the Hall resistivity $\rho_{xy}$ at various temperatures below $T_c$. (b) Magnetic field dependence of the Hall resistivity $\rho_{xy}$ at various temperatures above $T_c$ showing linear slopes. (c) Carrier density extrapolated from the slop of the Hall resistivity as a function of temperature. Negative numbers are used to indicate that the carriers remain to be electrons for the whole temperature range. (d) Temperature dependence of anomalous Hall conductivity $\sigma^{a}_{xy}$ in zero magnetic field. (e) $\sigma^{a}_{xy}$ as a function of $\sigma_{xx}$ and the fit to $\sigma^a_{xy} = a\sigma^2_{xx} + b/\sigma_{xx} + c$ to extract the intrinsic contribution. (f) $\rho^{a}_{xy}$ as a function of $\rho_{xx}$ and the fit to $\rho^a_{xy} = -a - b\rho_{xx}^3 - c\rho_{xx}^2$ to extract the intrinsic contribution.}  
\end{figure*}

\begin{figure*}[ht!]
    \includegraphics[width=1\linewidth]{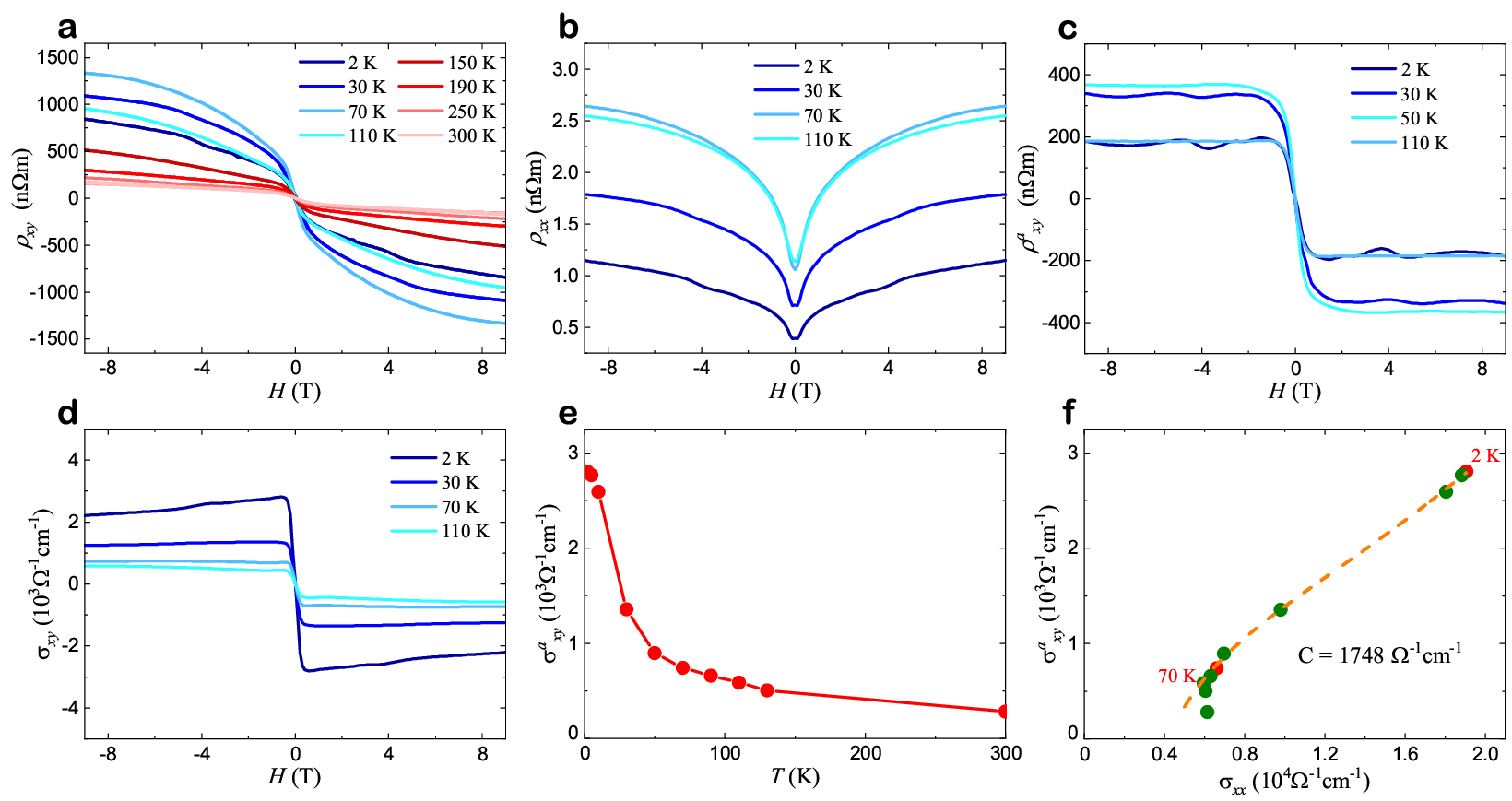}
    \caption{Anomalous Hall effect of sample S3. Magnetic field is applied along $c$ axis and electric current is along $a$ axis. (a) Magnetic field dependence of the Hall resistivity $\rho_{xy}$ at various temperatures. (b) Magnetic field dependence of the longitudinal resistivity $\rho_{xx}$ at various temperatures. (c) The anomalous Hall resistivity after subtracting the normal Hall resistivity obtained from two band Hall model. (d) Magnetic field dependence of the Hall conductivity $\sigma_{xy} = -\rho_{xy}/(\rho^2_{xy}+\rho^2_{xx})$ at various temperatures. (e) Temperature dependence of anomalous Hall conductivity $\sigma^{a}_{xy}$ in zero magnetic field. (f) $\sigma^{a}_{xy}$ as a function of $\sigma_{xx}$ and the fit to $\sigma^a_{xy} = a\sigma^2_{xx} + b/\sigma_{xx} + c$ to extract the intrinsic contribution.}  
\end{figure*}

\begin{figure}[t!]
    \includegraphics[width=1\linewidth]{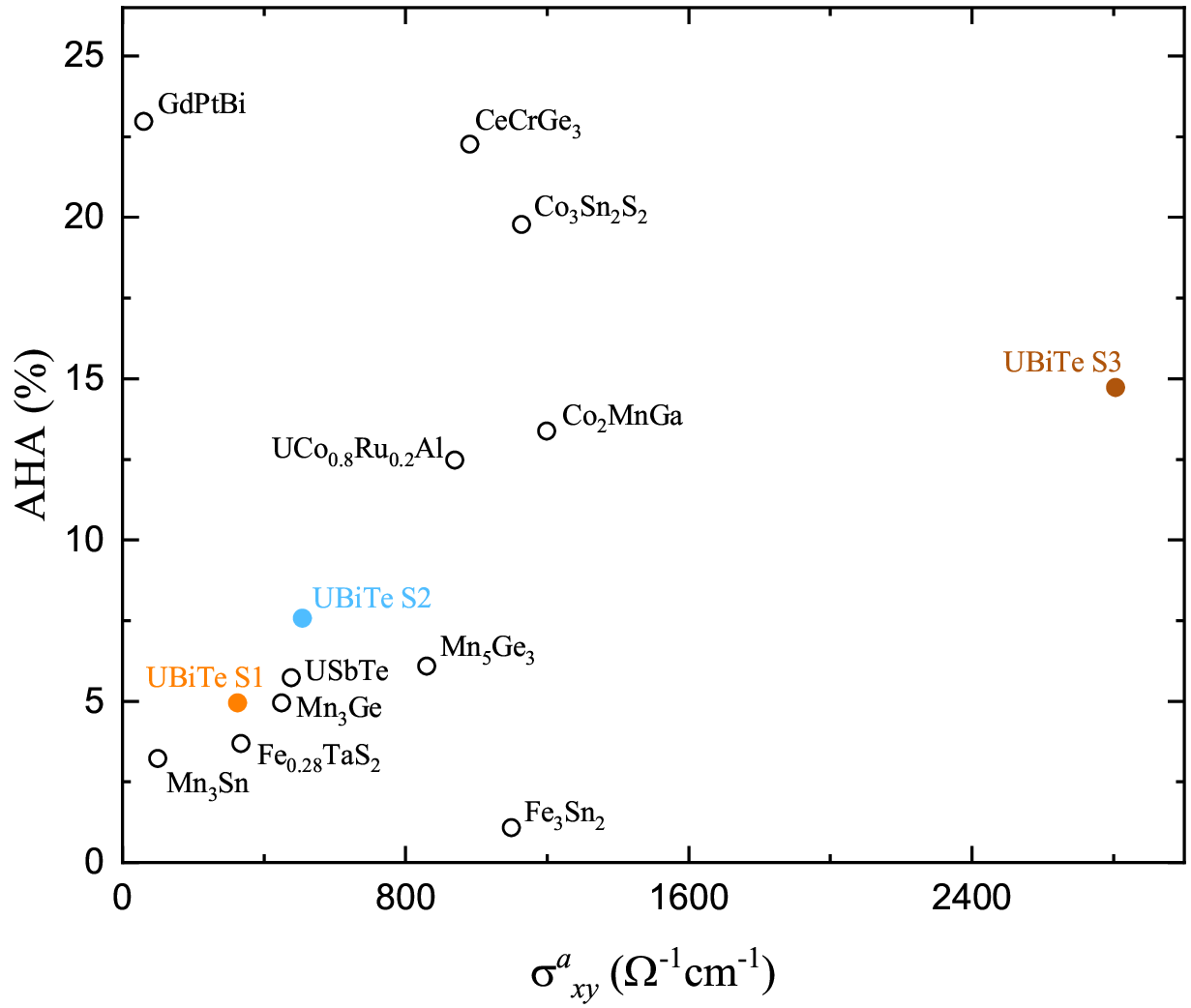}
    \caption{Anomalous Hall angle as a function of anomalous Hall conductivity in various topological magnetic materials. Data from literature are adopted from reference~\citenum{Li2024}.} 
\end{figure}

Figure 3 shows the Hall data for sample S2. In sharp contrast to sample S1, the Hall resistivity above $T_{c}$ exhibits a clear slope change in the magnetic field, indicating the presence of multiple types of carriers. As the temperature decreases, this feature gradually transitions into a linear slope, suggesting a shift to a single type of carrier. To make a direct comparison with S1, we use the single-band Hall model to fit the normal Hall resistivity below $T_{c}$ to extract the carrier density, as shown in Fig. 3c. To facilitate the comparison, we include the 70-100~K range, using the Hall resistivity above 1~T for the single-band model. The overall temperature dependence is similar to that of S1, with a peak around 70~K. However, the carrier density is a few orders of magnitude lower than that of S1, indicating that this sample is at a different Fermi level.

The anomalous Hall signal in this sample also displays notable differences when compared to S1. The hysteresis loop signifying the anomalous Hall effect has a different temperature dependence. In this sample, the hysteresis loop is barely visible at 50 K and completely disappears by 70 K, whereas in S1, the loop persists up to 70 K. Additionally, while the overall value of the AHC is similar to that of S1, the intrinsic contributions extracted from fitting both $\sigma^a_{xy}$ and $\rho^a_{xy}$ measurements are significantly smaller, yielding values of 420/$\Omega$-cm from $\sigma^a_{xy}$ and 545/$\Omega$-cm from $\rho^a_{xy}$. Similar to S1, the fitting remains valid only up to the Kondo coherence temperature, underscoring the correlation between Berry curvature and the hybridized bands. 

Figure 4 shows the Hall data for sample S3. Across the entire temperature range, $\rho_{xy}$ has an S shape instead of a linear slope, indicating that the normal Hall effect originates from multiple band carriers. However, a simple two-band Hall model does not fit $\rho_{xy}$. To fit $\rho_{xy}$ accurately, an additional Sigmoid function term is required. This indicates that an anomalous Hall signal is present, even though a clear rectangular hysteresis loop is not observed.


The anomalous Hall resistivity, extracted from a two-band Hall model fitting, is shown in Fig. 4c, yielding an exceptionally high AHC of 3000/$\Omega$-cm. However, the reliability of this value may be compromised by its dependence on the normal Hall model, which assumes only two types of carriers. Given the possibility that the normal Hall effect could arise from more complex carriers, we directly converted the $\rho_{xy}$ data to $\sigma_{xy} = -\rho_{xy}/(\rho^2_{xy}+\rho^2_{xx})$ using magnetoresistance data to reduce model dependency. This conversion revealed a distinct step-like feature in $\sigma_{xy}$, indicative of anomalous Hall behavior rather than multiple carrier effects. Although we do not observe a hysteresis loop, the step-like feature is more pronounced than what is typically observed in many other systems exhibiting anomalous Hall effects~\cite{Ye2018,Li2020,Yang2020,Kumar2021}. At base temperature, the AHC in this sample reaches 2700/$\Omega$-cm, one of the highest recorded for topological magnetic materials, as shown in Fig.~5. Using the same fitting function outlined above, we separated intrinsic and extrinsic contributions, yielding intrinsic values of 1970/$\Omega$-cm from $\rho^a_{xy}$ and 1750/$\Omega$-cm from $\sigma^a_{xy}$. These values are comparable to the maximum AHC value of USbTe from our previous DFT+U calculations~\cite{Siddiquee2023}.


\begin{table}[h]
\caption{SEM-EDX composition analysis}
\begin{tabular}{ l cc c cc}
 \toprule
  & Atomic & Composition & Percentage\\
Element  & S1 & S2 & S3 & \\
 \colrule
     U &  $35.7 \% \pm 1.8 \%$ & $35.4\% \pm 0.7\%$  & $ 35.6 \% \pm 0.8 \% $\\
     Bi & $32.7 \% \pm 1.0 \%$ & $31.7\% \pm 1.0\%$  & $ 30.9 \% \pm 0.5 \% $\\
     Te &  $31.6 \% \pm 1.6 \%$ & $32.9\% \pm 0.7\%$  & $ 33.6 \% \pm 1.0 \% $\\
\botrule
\end{tabular}
\label{stable_edx}
\end{table}

All three samples were synthesized under the same conditions and exhibit similar magnetic properties, as shown in the Supplementary Material. To understand the variations among samples, we performed a chemical composition analysis using Energy-Dispersive X-ray Spectroscopy (EDX) measurements. These measurements were taken in six different areas for each sample, and the average chemical composition and standard deviation are shown in Table 1. While there is some variation across the three samples, it is not very large, with a standard deviation of about 1$\%$. The uranium and tellurium percentages remain nearly identical, while there is a slight difference for Bi even considering the error bars. This small variation in Bi concentration can lead to slight modifications in the electronic structure of the material, particularly affecting the Fermi energy. These changes are consistent with the observed variations in normal Hall resistivity, highlighting the sensitivity of the electronic properties to compositional differences. Future experiments are required to further verity this hypothesis.


In concluding this study, we have demonstrated a number of critical observations across three samples of UBiTe: 1. There are very minimal variations in chemical composition. 2. Dramatic differences in carrier properties were observed, ranging from a single-band electron carrier with a density of 10$^{28}$/m$^3$ in Sample 1, to 10$^{22}$/m$^3$ in Sample 2, to distinctly multi-band carriers in Sample 3. These variations indicate different Fermi levels across the samples. 3. Our model for extracting the intrinsic contribution to the AHC is effective only up to the Kondo coherence temperature, highlighting a strong correlation between the intrinsic AHC and the Kondo hybridized bands. 4. The intrinsic contribution to AHC also shows considerable variation among the samples. Given the change in Fermi level, this suggests that the Berry curvature is predominantly hosted by the narrow Kondo hybridized bands that are very close to the Fermi level. In this case, even minor shifts in the Fermi level could lead to significant changes in hybridization, and consequently, affect both carrier density and Berry curvature, thereby dramatically influencing the material’s physical properties.

These distinctive features contrast sharply with behaviors observed in two related systems. In weakly correlated Weyl semimetals, the Berry curvature is associated with bands that exhibit sharp dispersion and low effective mass. In these systems, slight changes in the Fermi level do not lead to significant changes in physical properties. Meanwhile, in Kagome lattice systems, although strong correlation effects arise from flat bands, the topological characteristics of the system are primarily associated with Dirac cones, which are separate from the flat bands~\cite{Bolens2019}. In contrast, in UBiTe and similar Kondo lattice topological systems, the nontrivial topology is directly hosted by the flat Kondo bands, which integrate strong correlations with topological phenomena. Our study advances the understanding of the intrinsic link between Berry curvature and Kondo hybridization and illustrates the profound impact of Fermi level variations on the topological and electronic properties of Kondo lattice systems, distinguishing them from other types of topological materials.

\section{Acknowledgement}
We would like to acknowledge Jiun-Haw Chu for bringing reference~\citenum{Jones2022} to our attention. The work at Washington University is supported by the National Science Foundation (NSF) Division of Materials Research Award DMR-2236528. The work at the University of Arizona is supported by the NSF under Award No. DMR-2338229.  

\bibliography{UXTe}

\end{document}